\documentclass[reprint,10pt,aps,physrev,superscriptaddress,floatfix,twoside,nofootinbib]{revtex4-2}
\pdfoutput=1
\usepackage[english]{babel}
\usepackage[utf8]{inputenc}
\usepackage{amssymb,amsbsy,amsmath}
\usepackage{graphicx}
\usepackage{dcolumn}
\usepackage{bm}
\usepackage{bbold}
\usepackage{bbm}
\usepackage{subfigure}
\usepackage{braket}
\usepackage{textcomp}
\usepackage{placeins}

\usepackage{scalerel}
\usepackage{tikz}
\usetikzlibrary{calc}
\usetikzlibrary{patterns}
\usetikzlibrary{svg.path}
\definecolor{orcidlogocol}{HTML}{A6CE39}
\tikzset{
  orcidlogo/.pic={
    \fill[orcidlogocol] 
	svg{M256,128c0,70.7-57.3,128-128,128C57.3,256,0,198.7,0,128C0,57.3,57.3,0,128,0C198.7,0,256,57.3,256,128z};
    \fill[white] 
	svg{M86.3,186.2H70.9V79.1h15.4v48.4V186.2z}
	svg{M108.9,79.1h41.6c39.6,0,57,28.3,57,53.6c0,27.5-21.5,53.6-56.8,53.6h-41.8V79.1z M124.3,172.4h24.5c34.9,0,42.9-26.5,42.9-39.7c0-21.5-13.7-39.7-43.7-39.7h-23.7V172.4z}
    svg{M88.7,56.8c0,5.5-4.5,10.1-10.1,10.1c-5.6,0-10.1-4.6-10.1-10.1c0-5.6,4.5-10.1,10.1-10.1C84.2,46.7,88.7,51.3,88.7,56.8z};
  }
}

\newcommand\orcid[1]{\href{https://orcid.org/#1}{\mbox{\scalerel*{
\begin{tikzpicture}[yscale=-1,transform shape]
\pic{orcidlogo};
\end{tikzpicture}
}{|}}}}

\definecolor{midnight1}{HTML}{010c1e}
\definecolor{midnight2}{HTML}{001e38}
\definecolor{midnight3}{HTML}{4a6d88}
\definecolor{midnight4}{HTML}{c6cdd7}
\definecolor{ocean}{HTML}{004080}
\definecolor{aqua}{HTML}{0080ff}
\definecolor{altblue}{HTML}{16487a}

\usepackage[
	colorlinks,
	linkcolor=midnight3,
	citecolor=midnight3,
	urlcolor=midnight3,
	hypertexnames=false
	]{hyperref}
\usepackage{xurl}
\hypersetup{breaklinks=true}

\newcommand{\thubbard}{t_{\mathrm{h}}}

\newcommand{\hc}{\mathrm{h.c.}}
\newcommand{\inice}{\mathrm{i}}

\newcommand{\figref}[1]{Fig.~\protect\ref{#1}}

\newcommand{\Tr}[1]{\mbox{Tr}\left[\,{#1}\,\right]}

\newcommand{\hamiltonian}{\mathcal{H}}
\newcommand{\commutator}[2]{\left[\,#1,#2\,\right]}
\newcommand{\anticommutator}[2]{\left\lbrace\,#1,#2\,\right\rbrace}
\newcommand{\anylocaldensity}{\varrho}
\newcommand{\partitionfunction}{\mathcal{Z}}
\newcommand{\kboltzmann}{k_{{B}}}
\newcommand{\etothe}[1]{\mathrm{e}^{#1}}

\newcommand{\localdensityexp}[1]{q_{#1}(t)}

\newcommand{\aref}[1]{App.~\ref{#1}}
\newcommand{\sref}[1]{Sec.~\ref{#1}}

\definecolor{palette1}{HTML}{A8216B}
\definecolor{palette2}{HTML}{F1184C}
\definecolor{palette3}{HTML}{F36943}
\definecolor{palette4}{HTML}{F7DC66}
\definecolor{palette5}{HTML}{2E9599}

\begin{document}
\title[Spatiotemporal dynamics of classical and quantum density profiles]{Spatiotemporal dynamics of classical and
	quantum density profiles in low-dimensional spin systems}

\author{Tjark Heitmann \orcid{0000-0001-7728-0133}}
\email{tjark.heitmann@uos.de}
\affiliation{Department of Physics,
	University of Osnabrück, D-49076 Osnabrück, Germany}

\author{Jonas Richter \orcid{0000-0003-2184-5275}}
\affiliation{Department of Physics and Astronomy,
	University College London, Gower Street, London WC1E 6BT, UK}

\author{Fengping Jin \orcid{0000-0003-3476-524X}}
\affiliation{Institute for Advanced Simulation,
	Jülich Supercomputing Centre, Forschungszentrum Jülich, D-52425 Jülich, Germany}

\author{Kristel Michielsen \orcid{0000-0003-1444-4262}}
\affiliation{Institute for Advanced Simulation,
	Jülich Supercomputing Centre, Forschungszentrum Jülich, D-52425 Jülich, Germany}

\author{Hans De Raedt \orcid{0000-0001-8461-4015}}
\affiliation{Zernike Institute for Advanced Materials,
	University of Groningen, NL-9747 AG Groningen, Netherlands}

\author{Robin Steinigeweg \orcid{0000-0003-0608-0884}}
\affiliation{Department of Physics,
	University of Osnabrück, D-49076 Osnabrück, Germany}

\date{\today}

\begin{abstract}
	We provide a detailed comparison between the
	dynamics of high-temperature spatiotemporal correlation functions in quantum
	and classical spin models. In the quantum case, our
	large-scale numerics are based on the concept of quantum typicality,
	which exploits the fact that random pure quantum states can faithfully
	approximate ensemble averages, allowing the simulation of spin-$1/2$ systems
	with up to
	$40$ lattice sites.
	Due to the exponentially growing Hilbert space, we find that for such system
	sizes even a single random state is sufficient to yield results
	with extremely low noise that is negligible for most practical purposes.
	In contrast, a classical analog of typicality is missing.
	In particular, we
	demonstrate that, in order to obtain data with a similar level of noise
	in the classical case,
	extensive averaging over classical trajectories is required,
	no matter how large the system size.
	Focusing on (quasi-)one-dimensional
	spin chains and ladders, we find a remarkably good agreement between
	quantum and classical
	dynamics. This
	applies not only to cases where both the quantum and classical model
	are nonintegrable, but also to cases where the quantum spin-$1/2$
	model is
	integrable and the corresponding classical $s\to\infty$ model is not.
	Our analysis is based on the comparison
	of space-time profiles of the spin and
	energy correlation functions, where the agreement is found to hold not only
	in the bulk but also in the tails of the resulting density
	distribution. The
	mean-squared displacement of the
	density profiles reflects the nature of emerging
	hydrodynamics and is found to exhibit
	similar scaling for quantum and classical
	models.
\end{abstract}

\maketitle

\section{Introduction}
Building on seminal results in chaos and random-matrix theory
\cite{Brody1981,Bohigas1984} as well
as more recent developments such as the eigenstate thermalization hypothesis
\cite{Deutsch1991,Srednicki1994,Rigol2008,Dalessio2016},
it is
now well established that generic quantum and classical many-body systems
relax to thermal equilibrium at long times
\cite{Polkovnikov2011,Eisert2015,Gogolin2016,Dalessio2016,Borgonovi2016,
	Bertini2021}.
In this context, one of the most generic nonequilibrium situations
is given by transport processes of local densities due to a global conservation
law \cite{Bertini2021}. Such transport processes
describe the slow relaxation from local to global equilibrium
and dominate the late-time and long-wavelength properties
of systems with conservation laws.
Gaining a deeper understanding how such a macroscopic hydrodynamic behavior
emerges from the underlying microscopic equations of motion is the subject of
ongoing theoretical research
\cite{Bertini2021, Khemani2018},
while novel
experiments with different quantum-simulator
platforms nowadays allow the controlled
exploration even of anomalous types of
quantum transport and hydrodynamics \cite{Jepsen2020, Wei2022, Joshi2022}.
Key insights have been gained not least due to
improved numerical machinery
\cite{Wurtz2018, Paeckel2019, Rakovszky2022, Ye2020, Heitmann2020,
	White2018},
as well as the introduction of suitable
random-circuit models, which provide minimal models to capture the universal
properties of chaotic quantum systems
\cite{Nahum2017,VonKeyserlingk2018,Nahum2018,Khemani2018,Moudgalya2021,
	Bertini2019}.

An important role, which can strongly influence the nature of transport in a
given model, is played by integrability. On one hand, in the case
of classical mechanics, integrability is well defined in terms of
the Liouville-Arnold theorem, which requires $L$ (mutually commuting)
constants of motion for a
system of $L$ spins \cite{Arnold1978,Steinigeweg2009}.
The trajectories in such integrable systems remain
strictly confined to a small part of phase space, resulting in a breakdown of
ergodicity. In contrast, if there are not enough constants of motion,
integrability is absent and chaotic dynamics is expected to emerge as a
consequence. On the other hand, such a clear-cut definition of integrability is
not available in the case of quantum systems \cite{Caux2011}.
One commonly applied definition
is solvability in terms of the Bethe Ansatz, which
includes important systems such as the spin-$1/2$ Heisenberg chain and the
one-dimensional Fermi-Hubbard model
\cite{Bethe1931,Korepin1993,Takahashi1999,Levkovich-Maslyuk2016,Essler2005}.
In particular, for such models it is
possible to construct extensive sets
of (quasi)local integrals of motion \cite{Prosen2011,Ilievski2016,Bertini2021},
reminiscent of the definition of integrability in classical mechanics.
Building on this intricate algebraic structure, much progress in
understanding the dynamics of integrable quantum systems has been recently made
within the framework of generalized
hydrodynamics \cite{Bertini2016,Castro-Alvaredo2016,Doyon2020},
which provides analytical
support for early numerical studies \cite{Bertini2021}. In particular, while
integrable
systems (due to their coherent quasiparticle excitations) are often expected to
exhibit ballistic transport and finite Drude
weights \cite{Bertini2021}, the latter indicating
that induced currents remain (at least partially)
conserved on indefinite time scales, it has become clear that
the dynamics of integrable systems is much richer.
This includes parameter regimes where normal
diffusive transport \cite{Znidaric2011, Steinigeweg2017}, usually associated
with chaos, can occur.
Moreover, if
the integrable model additionally
possesses a non-Abelian
symmetry, it was found
that transport is neither ballistic nor
diffusive, but superdiffusive instead \cite{Ilievski2021,Bulchandani2021}.
More specifically,
it has been argued that this combination of
integrability and non-Abelian
symmetry generically leads to
superdiffusion within the Kardar-Parisi-Zhang
(KPZ) universality class with dynamical exponent
$z = 3/2$
\cite{Kardar1986,Ilievski2018,Gopalakrishnan2019a,Ljubotina2019,Das2019,Dupont2020,
	DeNardis2020,Weiner2020,Bertini2021,Bulchandani2021},
first reported numerically in the case of high-temperature spin
transport in the
spin-$1/2$ Heisenberg chain
which is SU$(2)$ symmetric \cite{Ljubotina2019,Ljubotina2017}.

Integrability
only accounts for a small region of
the full parameter space and is typically
found in one-dimensional models.
Especially in the case of quantum many-body
systems, it appears that integrability
is very fragile and even
a tiny integrability-breaking perturbation
will induce chaos and generic behavior in the
limit of large system sizes and
long times \cite{Dalessio2016}. Nevertheless,
interesting phenomena such as
prethermalization can
occur in the regime close to integrability \cite{Reimann2019a,Mallayya2019},
and it is an interesting
question
to what extent the dynamics of weakly perturbed systems
can be understood due to their vicinity to an
integrable point
\cite{Heitmann2021,Dabelow2020,Richter2020a,Bastianello2021,Claeys2021}.
Generally, however, the
vast majority of quantum and classical systems,
particularly in spatial
dimensions larger than one, are nonintegrable and generic.
For such systems, one
typically expects that the emerging transport behavior at long times is given
by
standard diffusion. Such normal diffusive
transport has indeed been numerically
confirmed in a variety of models
\cite{Steinigeweg2017,Richter2019f,Lux2014,Bohrdt2017,Richter2018a,
	Richter2019g,Mcroberts2021}.
However,
the numerical extraction of quantitative values for transport coefficients
(e.g.,
diffusion constant), especially in the quantum case, is quite challenging and
still an actively pursued direction
of research \cite{Rakovszky2022, Ye2020, White2018}.
Furthermore, even in the case of
nonintegrable systems which fulfill various indicators of quantum and classical
chaos, there still exist
counterexamples to the expected
diffusive behavior. This includes models with
additional symmetries or conservation
laws and models with kinetic constraints
\cite{Moudgalya2021,Richter2022a,Singh2021},
which can host anomalous types of transport such as subdiffusion, as well as
long-range systems where transport
can become superdiffusive \cite{Kloss2019,Schuckert2020,Richter2022}.

Even though strongly correlated many-body quantum systems generally do not
have an obvious classical limit, a natural choice in the case of quantum spin
models is to consider the limit of infinite spin quantum number $s \to
	\infty$, where the quantum spin operators become classical
three-dimensional vectors.
In this context, it is an intriguing
question to ask whether and to what extent the dynamics, and in particular the
transport properties, of the quantum and the corresponding classical spin model
agree with each other. In particular, even though the emerging
late-time transport behavior of quantum models
may effectively be described by a
classical hydrodynamic theory (e.g., a diffusion equation),
it is not obvious that quantum
and classical dynamics agree on a
detailed quantitative level (for instance
regarding the explicit value of
diffusion coefficients). Such a
quantitative agreement is nontrivial even
at high temperatures, where quantum
effects are less pronounced, due to the
different microscopic equations of
motion. Moreover, sending $s \to \infty$ can break the integrability of the
original spin-$1/2$ model. Complementing earlier work in
this direction \cite{Richter2020,Schubert2021,Starkov2018}, the goal of this
paper is
to provide a comprehensive
comparison between quantum and classical dynamics in
models of interacting spins. To this end, we focus on the
buildup of spatiotemporal correlation functions of local spin and energy
densities, which probe transport properties in the linear response
regime \cite{Bertini2021}, and are also intimately related to experimentally
accessible
quantities such as the spin structure factor measurable with
inelastic neutron scattering \cite{Lake2013}.

Studying the dynamics of quantum many-body systems is notoriously challenging
due to the exponentially growing Hilbert space. In this paper, we
rely on the concept of quantum typicality \cite{Heitmann2020,Jin2021},
which refers to the fact
that even a pure random quantum state can faithfully approximate the full
ensemble average. In particular, as we will explain below in more detail, the
statistical error of quantum typicality decreases exponentially with the size
of the system. As a consequence, significantly less averaging over random
states is required in larger systems
to obtain the same accuracy. Combined with efficient sparse-matrix techniques
for the time evolution of pure quantum states,
quantum typicality enables us to
simulate spatiotemporal correlation functions in
systems with Hilbert-space
dimensions far beyond the range of full
exact diagonalization. More specifically, we solve
the time-dependent Schr\"odinger equation for
models with up to $40$ spin-$1/2$ degrees of freedom, i.e.,
the total Hilbert space has
dimension $2^{40} \approx 10^{12}$, which
yields converged results for the buildup of spatiotemporal correlation
functions on sufficiently long time scales to extract the asymptotic
hydrodynamic behavior. Crucially, we demonstrate that for such enormous
Hilbert-space dimensions, the statistical fluctuations of
quantum typicality are strongly suppressed, such
that even a single random state approximates
the spatiotemporal correlation function with an
extremely low level of noise that is negligible
for all practical purposes. These results are
then compared to the corresponding classical system.
In contrast to quantum systems, the phase space
of classical mechanics only grows linearly with the number of lattice spins,
such that simulations are significantly less costly and much larger system
sizes
can be treated in principle.
However, as we demonstrate in this paper, a
classical analog of the concept of
typicality is missing. In particular, we show
that the statistical fluctuations in the classical trajectories are not
reduced with increasing system size, such that it remains
necessary even for larger and larger systems
to perform extensive statistical averaging over a high number of
trajectories to achieve the same low noise level as in the quantum case.

Focusing on (quasi-)one-dimensional spin chains and ladders, we
typically find a remarkably good agreement between quantum and classical
dynamics. This
applies not only to cases where both the quantum and classical model
are nonintegrable, but also to cases where the quantum model is
integrable and the corresponding classical model is not.
Our analysis is based on the comparison
of space-time profiles of the spin and
energy correlation functions, where the agreement is found to hold not only
in the bulk but also in the tails of the resulting density
distribution. This fact also manifests itself in
the time dependence of the
mean-squared displacement of the
density profiles, which reflects the nature of emerging
hydrodynamics and exhibits very similar scaling for quantum and classical
models, at least on the
time and length
scales considered here, with the exception of
cases where transport is dominated by integrability.
Furthermore, we show
that such a correspondence between quantum and classical
dynamics can also be achieved in less obvious cases where the
original quantum system is not directly written in spin language. In
particular, we consider the one-dimensional Fermi-Hubbard model, which by means
of a Jordan-Wigner transform can be brought into the form of a particular type
of spin ladder, for which we then take the $s\to \infty$ limit.

The rest of this paper is structured as follows. In \sref{sec:model}, we
introduce the quantum spin models, their classical counterparts, as well as the
corresponding observables which are studied in this paper. Our numerical
approach based on quantum typicality is introduced in \sref{sec:methods},
where we also explain the methods used to integrate the quantum and classical
equations of motion as well as the role of averaging.
We present our results for spin and energy transport in \sref{sec:results}, where we consider spin chains in
\sref{sec:results-XXZ-chain} and spin
ladders in \sref{sec:results-XXX-ladder}.
Moreover, we discuss charge transport in the Fermi-Hubbard chain in
\sref{sec:results-hubbard}.
We summarize our findings and conclude in \sref{sec:conclusion}.

\section{Models and observables}\label{sec:model}

\subsection{Models}
In this paper, we consider different versions of {(quasi-)one}-dimensional
lattice models described by Hamiltonians of the form
\begin{align}\label{eq:LatticeHamiltonian}
	\hamiltonian=\sum\limits_{r=1}^{L}h_r
\end{align}
with periodic boundary conditions $L+1\equiv1$.
For quantum spin models, the lattice sites are occupied by stationary
spins with spin quantum number $s$,
represented by spin vector operators
$\mathbf{s}_{r}=(s_r^x,s_r^y,s_r^z)$. Their components fulfill the
defining spin algebra $(\hbar=1)$,
\begin{align}\label{eq:spin-algebra}
	\commutator{s_r^{\mu}}{s_{r'}^{\nu}}=
	\inice\,\delta_{rr'}\,\varepsilon_{\mu\nu\lambda}\,s_r^{\lambda}\ ,
\end{align}
where $\delta_{rr'}$ is the Kronecker delta,
$\varepsilon_{\mu\nu\lambda}$ is the antisymmetric Levi-Civita symbol,
and $\mu,\nu,\lambda\in\lbrace x,y,z\rbrace$. For spin quantum number
$s=1/2$, the components can be expressed in terms of Pauli matrices,
$s_r^{\mu}=\sigma_r^{\mu}/2$.

First, we consider the anisotropic Heisenberg chain (XXZ chain) with local
energy terms
\begin{align}\label{eq:local-hamilton-spinchain}
	h_r=J\left(s_r^xs_{r+1}^x+s_r^ys_{r+1}^y
	+ \Delta s_r^zs_{r+1}^z\right)\ ,
\end{align}
where $J>0$ is the antiferromagnetic exchange coupling
constant and $\Delta$ parametrizes the anisotropy in $z$ direction. For
any anisotropy, the total magnetization $S^z=\sum_rs_r^z$ is conserved,
$\commutator{\hamiltonian}{S^z}=0$.
We note that the spin-$1/2$ XXZ chain is integrable in terms of the Bethe Ansatz, which has
consequences for the transport properties of the model.
In
particular, the energy current is an exact constant of
motion such that energy transport in the spin-$1/2$ XXZ chain is dissipationless
for all values of $\Delta$ \cite{Bertini2021}.
Therefore, we here focus on the dynamics of
magnetization which can exhibit various types of behavior depending on the
choice of $\Delta$, as discussed in detail in
\sref{sec:results-XXZ-chain} below.
On the other hand, when considering the
classical version of the XXZ chain with $s \to \infty$
(cf.\ \sref{Sec::Classical}), the integrability of the model is
broken such that
one would naively expect chaotic dynamics resulting in the emergence
of diffusive transport. While diffusive energy transport has indeed been found
in classical XXZ chains, it turns out that observing clean spin
diffusion in all $\Delta$ regimes is a subtle issue
\cite{Muller1988,Gerling1989,Gerling1990,DeAlcantaraBonfim1992,
	DeAlcantaraBonfim1993,Bohm1993,Srivastava1994,Li2019,Bagchi2013}. This might be
related to the fact that taking the classical limit $s \to \infty$ is in some
sense only a ``weak'' integrability-breaking perturbation
\cite{Roy2022,DeNardis2021,Claeys2021}, as it leaves
the overall structure (such as the symmetries) of the Hamiltonian intact. As a
consequence, the impact of this perturbation on the original quantum dynamics
might be less pronounced.
As we demonstrate in \aref{app:longXXZ}, observing the onset of standard spin
diffusion in classical
spin chains (in a $\Delta$ regime where
quantum dynamics is ballistic) is indeed extremely challenging and
requires the analysis of large system sizes on long time scales.

Second, as a quasi-1D spin model, we study the isotropic Heisenberg
ladder (XXX ladder),
\begin{align}\label{eq:local-hamilton-spinladder}
	h_r=J\sum_{l=1,2}\mathbf{s}_{r,l}\cdot\mathbf{s}_{r+1,l}+
	\frac{J}{2}\sum_{r'=r}^{r+1}\mathbf{s}_{r',1}\cdot\mathbf{s}_{r',2}\ ,
\end{align}
where the local energy is defined on a ``plaquette''
consisting of four spins.
As above, the total magnetization
$S^z=\sum_{r,l}s^z_{r,l}$
is conserved. However, in contrast to the XXZ chain, the XXX
ladder is nonintegrable for $s = 1/2$. Thus, this is an example where both the
quantum and the classical model are nonintegrable.

Moving deeper into the realm of ``genuinely quantum'' models, we also
consider the Fermi-Hubbard chain with local Hamiltonians
\begin{align}\label{eq:local-hamilton-hubbard}
	h_r=- &\thubbard\sum_{\sigma=\uparrow,\downarrow}
	    \left(c_{r\vphantom{l},\sigma}^{\dag}
	c_{r+1,\sigma}^{\vphantom\dag}+\hc\right)            \\\notag
	+\, & U\left(n_{r,\uparrow}-\frac{1}{2}\right)
	\left(n_{r,\downarrow}-\frac{1}{2}\right)\ ,
\end{align}
where $\thubbard$ is the hopping amplitude of the spin-$\sigma$
fermions and $U$ is the on-site interaction strength. The creation
operator $c_{r,\sigma}^{\dag}$ creates a spin-$\sigma$ particle at site
$r$, whereas the annihilation operator $c_{r,\sigma}^{\phantom\dag}$
annihilates a spin-$\sigma$ particle at site $r$.
They fulfill the fermionic anticommutation relations,
\begin{align}
	\anticommutator{c_{r,\sigma}^{\phantom\dag}}{c_{r',\sigma}^{\phantom\dag}}=0
	\ ,
	\quad
	\anticommutator{c_{r,\sigma}^{}}{c_{r',\sigma}^{\dag}}=\delta_{rr'}\ ,
\end{align}
and define the local particle number operator
${n_{r,\sigma}=c_{r,\sigma}^{\dag}c_{r,\sigma}^{\phantom\dag}}$.
Similar to the XXZ chain, the Fermi-Hubbard chain is a prime
example of an integrable quantum system. In particular, as in the XXZ chain,
energy transport in the Fermi-Hubbard chain is ballistic for all values of $U$
such that we here focus only on charge transport.
Notably, by Jordan-Wigner
transformation \cite{Jordan1928},
this model is in turn equivalent to a
modified version of the spin ladder,
\begin{align}\label{eq:local-hamilton-hubbard-to-heisenberg}
	h_r=- & 2J_{\parallel}\sum_{l=1,2}
	\left(s_{r\vphantom{l},l}^x s_{r+1,l}^x+
	s_{r\vphantom{l},l}^y s_{r+1,l}^y \right)                                 \\\notag
	+\,   & \frac{J_{\bot }}{2}\sum\limits_{r'=r}^{r+1}\,s_{r',1}^zs_{r',2}^z
\end{align}
with $J_{\parallel}=\thubbard$ and $J_{\bot}=U$. Each leg of the spin
ladder is associated with one of the fermionic species $\uparrow$ or
$\downarrow$. The local magnetizations in the spin formulation
correspond to occupation numbers in the Hubbard language,
\begin{align}
	n_{r,\sigma}=s_{r,l}^z+\frac{1}{2}\ .
\end{align}

\subsection{Setup and observables}
We study the dynamics of local densities $\anylocaldensity_r$ of either
magnetization ($M$) or energy ($E$) based on the time-dependent
density-density correlation function,
\begin{align}
	C_{r,r'}(t)=\braket{\anylocaldensity_r(t)\anylocaldensity_{r'}}\ ,
\end{align}
where $\braket{\bullet}=\Tr{\rho_{\beta}\bullet}$ denotes the
expectation value with respect to the canonical density matrix
$\rho_{\beta}=\exp(-\beta\hamiltonian)/\partitionfunction$ and
$\partitionfunction=\Tr{\exp(-\beta\hamiltonian)}$ is the canonical
partition function at inverse temperature $\beta=1/\kboltzmann T$.
Here, operators evolve in time according to the Heisenberg picture,
i.e., $\anylocaldensity_r(t)=\exp(\inice\hamiltonian t)\,
	\anylocaldensity_r\exp(-\inice\hamiltonian t)$.

In the following, we fix $r'=L/2$ and study the time dependence of the
profile
\begin{align}\label{eq:spatial-correlation}
	C_r(t)\equiv C_{r,L/2}(t)=
	\braket{\anylocaldensity_r(t)\anylocaldensity_{L/2}}\ .
\end{align}
We focus on the high-temperature limit $\beta\rightarrow0$, where
$\rho_{\beta}\rightarrow1/d^L$ with Hilbert-space
dimension ${d^L}$ ($d$ denotes the model-specific local Hilbert-space
dimension).
Accordingly, the density profile is obtained by calculating
\begin{align}\label{eq:infinite-temp-correlation}
	C_r(t)=
	\frac{\Tr{\anylocaldensity_r(t)\anylocaldensity_{L/2}}}{d^L}\ ,
\end{align}
where the different local densities, depending on the system's geometry,
are defined as
\begin{align}
	\anylocaldensity^{(M)}_r=
	\begin{cases}
		\ s^z_r               & , \quad\mathrm{XXZ\ chain}  \\
		\ s^z_{r,1}+s^z_{r,2} & , \quad\mathrm{XXX\ ladder}
	\end{cases}
\end{align}
and
\begin{align}
	\anylocaldensity^{(E)}_r=h_r\ .
\end{align}
Moreover, in the case of charge transport in the Fermi-Hubbard
chain, we have $\varrho_r = n_{r,\uparrow} + n_{r,\downarrow} - 1 = s_{r,1}^z +
	s_{r,2}^z$, i.e., analogous to the case of spin dynamics in the XXX ladder.

\begin{figure}[t]
	\centering
	\includegraphics[width=\columnwidth]{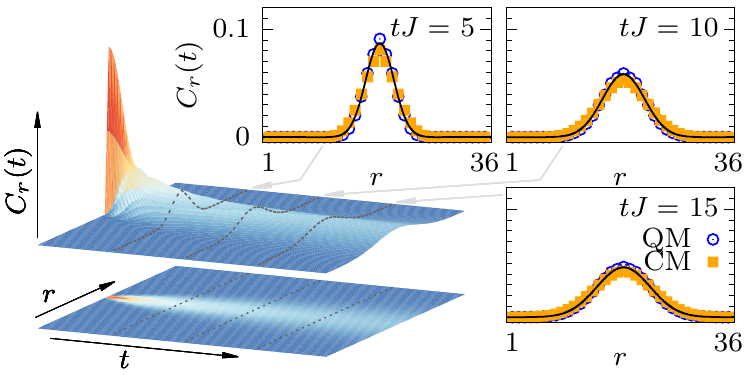}
	\caption{Exemplary plot of an initially peaked density profile that broadens over time by diffusion. The data shows quantum and classical
		magnetization dynamics in the XXZ spin chain with anisotropy $\Delta=1.5$ and system size $L=36$. For a more detailed discussion of the results,
		see \sref{sec:results-XXZ-chain} below. 
		Note that the legend in the lower right panel applies to all other panels.
		}
	\label{fig:QvC-3d}
\end{figure}

Initially, a peaked spatial distribution arises for the local
magnetizations,
\begin{align}\label{eq:initial-peak}
	C^{(M)}_r(t=0)
	\begin{cases}
		\ \neq 0 & , \quad r=L/2        \\
		\ = 0    & , \quad\mathrm{else}
	\end{cases}\ ,
\end{align}
as can be seen from the spatiotemporal density profiles shown
in \figref{fig:QvC-3d}.
A similarly peaked initial density distribution also arises
for $C_r^{(E)}(t=0)$,
albeit accompanied by two smaller peaks at adjacent lattice sites
$L/2\pm1$ due to shared bonds between local energy terms $h_r$ and
$h_{r\pm1}$.
The main contribution of
this work is to provide a detailed comparison between the real-time broadening
of such density profiles for quantum and classical spin models.

\subsection{Classical limit}\label{Sec::Classical}
The classical counterpart of the quantum spin models introduced above
is achieved by taking the limits of both $\hbar\rightarrow0$ and
$s\rightarrow\infty$ while maintaining the constraint
$\hbar\sqrt{s(s+1)}=\mathrm{const.}$
Then, the spin operators become three-dimensional vectors of constant
length, $\vert{\bf s}_r\vert=1$, and the
relation \eqref{eq:spin-algebra} turns into
\begin{align}
	\anticommutator{s_r^{\mu}}{s_{r'}^{\nu}}=
	\delta_{rr'}\,\varepsilon_{\mu\nu\lambda}\,s_r^{\lambda}\ ,
\end{align}
where $\anticommutator{\bullet}{\bullet}$ denotes the Poisson bracket.
Consequently, the time evolution of each spin is determined by the
Hamiltonian equations of motion,
\begin{align}\label{eq:hamiltonian-eom}
	\dot{\bf s}_r =
	\left\lbrace{\bf s}_r,\hamiltonian\right\rbrace =
	\frac{\partial \hamiltonian}{\partial {\bf s}_r}\times{\bf s}_r\ .
\end{align}
The infinite-temperature correlation function
\eqref{eq:infinite-temp-correlation} can be obtained in the classical
case by taking $\braket{\bullet}$ as an average over trajectories in
phase space,
\begin{align}\label{eq:classical-averaging}
	C_r(t)\approx\frac{1}{N}
	\sum_{n=1}^N\anylocaldensity_r(t)\anylocaldensity_{L/2}(0)\ .
\end{align}
For each of the $N\gg 1$ realizations, the initial configurations
$\mathbf{s}_r(0)$ are drawn at random.

\subsection{Diffusion on a lattice}\label{sec:lattice-diffusion}
The correlation functions $C_r(t)$ can be connected to the time dependence of
local densities $\localdensityexp{r}$ in a scenario, where
the initial state is prepared close to the canonical equilibrium density matrix
$\rho_{\beta}$ as
\begin{align}
	\rho(0)\propto\etothe{-\beta(\hamiltonian-\varepsilon\anylocaldensity_{L/2})}\ ,
\end{align}
which can be expanded in $\varepsilon$ and, for high temperatures, takes on the simple form
\begin{align}
	\rho(0)\propto\mathbb{1}+\beta\varepsilon\anylocaldensity_{L/2}\ .
\end{align}
For this initial state and using $\Tr{\anylocaldensity_r}=0$,
\begin{align}
	\localdensityexp{r}
	=\braket{\anylocaldensity_r(t)}
	 & =\Tr{\anylocaldensity_r(t)\rho(0)}                      \\\notag
	 & \propto\Tr{\anylocaldensity_r(t)\anylocaldensity_{L/2}}
	\propto C_r(t)\ ,
\end{align}
i.e., $C_r(t)$ describes the dynamics of the local densities $\localdensityexp{r}$ after
an initial density distribution of the form \eqref{eq:initial-peak}.
Speaking differently, $C_r(t)$ can be interpreted as the
dynamics and relaxation of some initial spin or energy excitation evolving on
top of a featureless infinite-temperature many-body background.

The local densities $\localdensityexp{r}$ show diffusive
transport, if they fulfill the lattice diffusion equation,
\begin{align}\label{eq:diffusion-equation}
	\frac{d}{dt}\localdensityexp{r}=D\left[\localdensityexp{r-1}-2\localdensityexp{r}+\localdensityexp{r+1}\right]
\end{align}
with some diffusion constant $D$. The temporal growth of the spatial
variance,
\begin{align}\label{eq:spatial-variance}
	\Sigma^2(t)=\sum\limits_{r=1}^Lr^2\delta \localdensityexp{r}
	-\left[\sum\limits_{r=1}^L r\delta \localdensityexp{r}\right]^2\ ,
\end{align}
with ${\delta \localdensityexp{r}\propto \localdensityexp{r}}$ normalized to
${\sum_r \delta \localdensityexp{r}=1}$ for all times $t$, can also be
used for characterizing the dynamics.
A scaling according to ${\Sigma(t)\propto t^{\alpha}}$ is called
	{\it ballistic} for ${\alpha = 1}$, {\it superdiffusive} for
${1/2 < \alpha < 1}$, {\it diffusive} for ${\alpha = 1/2}$,
{\it sub\-diffusive} for ${0< \alpha < 1/2}$, and {\it insulating}
for ${\alpha = 0}$.

Additionally, for initial density distributions of the form \eqref{eq:initial-peak},
the solution of the diffusion equation \eqref{eq:diffusion-equation} reads
\begin{align}
	\delta \localdensityexp{r}=\exp(-2Dt)I_{r-L/2}(2Dt)\ ,
\end{align}
where $I_{r}(t)$ is the modified Bessel function of first kind and of
order $r$.
The corresponding spatial dependence for fixed times $t$ is well
approximated by Gaussian functions,
\begin{align}\label{eq:gaussianfit}
	\delta \localdensityexp{r}=\frac{1}{\Sigma(t)\sqrt{2\pi}}
	\exp\left[-\frac{(r-L/2)^{2}}{2\Sigma^2(t)}\right]\ ,
\end{align}
where $\Sigma(t)=2Dt$.
While the scaling analysis of the spatial width \eqref{eq:spatial-variance}
may hint at the existence of diffusive transport, the form
\eqref{eq:gaussianfit} of the spatial dependence of the density distribution
is a precise diagnostics.

\section{Numerical methods}\label{sec:methods}
\subsection{Dynamical quantum typicality}
For the quantum systems, we employ the concept of dynamical quantum typicality (DQT)
\cite{LLoyd1988,Hams2000,Gemmer2004,Iitaka2004,Goldstein2006,Reimann2007,Bartsch2009,Reimann2018},
which
essentially allows us to replace the trace in the calculation of the
correlation function \eqref{eq:spatial-correlation} by a scalar product
between two auxiliary pure states \cite{Elsayed2013,Steinigeweg2014},
\begin{align}\label{eq:typicality-correlation-finite-temp}
	C_r(t)=\bra{\phi_{\beta}(t)}\anylocaldensity_r
	\ket{\varphi_{\beta}(t)}+\epsilon(\ket{\phi})\ ,
\end{align}
where the states
\begin{align}
	\ket{\varphi_{\beta}(t)}=\etothe{-\inice\hamiltonian t}
	\anylocaldensity_{L/2}\ket{\phi_{\beta}}
	\ ,\quad
	\ket{\phi_{\beta}(t)}=
	\etothe{-\inice\hamiltonian t}\ket{\phi_{\beta}}
\end{align}
are constructed with
\begin{align}
	\ket{\phi_{\beta}}=
	\frac{\sqrt{\rho_{\beta}}\ket{\phi}}{\sqrt{\bra{\phi}\rho_{\beta}\ket{\phi}}}\ .
\end{align}
The typical reference state $\ket{\phi}$ is constructed as a random superposition of states
$\ket{k}$ in the given orthonormal basis,
\begin{align}
	\ket{\phi}=\sum_{k=1}^{d^L}c_k\ket{k}\ ,
\end{align}
where the complex coefficients $c_k$ are randomly drawn from a
distribution which is invariant under all unitary transformations in
the Hilbert space (Haar measure) \cite{Bartsch2009}. In practice, the real and imaginary
parts of the $c_k$ are drawn independently from a  standard normal
distribution.
The variance of the statistical error $\epsilon(\ket{\phi})$ that arises in
Eq.~\eqref{eq:typicality-correlation-finite-temp} is bounded from above \cite{Jin2021},
\begin{align}\label{eq:typicality-error}
	\sigma(\epsilon)<
	\mathcal{O}\left(\frac{1}{\sqrt{\rule{0pt}{3mm}
			\dim_{\mathrm{eff}}}}\right)\ ,
\end{align}
where $\dim_{\mathrm{eff}}=\Tr{\etothe{-\beta(\hamiltonian-E_0)}}$
with ground-state energy $E_0$
is the effective Hilbert-space dimension at inverse temperature $\beta$.
In the infinite-temperature limit,
$\lim_{\beta\rightarrow0}\dim_{\mathrm{eff}}=d^L$, which renders the
typicality error negligibly small for the system sizes considered here.
Additionally, for $\beta\rightarrow0$, the calculation of
Eq.~\eqref{eq:typicality-correlation-finite-temp} can be further simplified to \cite{Richter2019c}
\begin{align}
	C_r(t)=\bra{\psi(t)}\anylocaldensity_r\ket{\psi(t)}+\epsilon(\ket{\phi})
\end{align}
using just one pure state
\begin{align}
	\ket{\psi(0)}=
	\frac{\sqrt{\anylocaldensity_{L/2}+c}\ket{\phi}}{\sqrt{\braket{\phi\vert\phi}}}\ ,
\end{align}
where the constant $c$ ensures that the operator $\anylocaldensity_{L/2}+c$
has nonnegative eigenvalues.

The time dependence is now a property of the pure states and can be obtained
by iteratively applying the time evolution in small time steps,
\begin{align}
	\ket{\psi(t+\delta t)}=
	\etothe{-\inice\hamiltonian \delta t}\ket{\psi(t)}\ ,\quad \delta t \ll J\ .
\end{align}
For each time step, the action of the time-evolution operator on the state is
obtained by massively parallelized simulations on supercomputers,
which rely on both Trotter decompositions
\cite{Suzuki1985,DeRaedt1987} and Chebyshev-polynomial
expansions \cite{Dobrovitski2003,Weisse2006}.

\subsection{Classical averaging}
\def\numtrajectories{300}
\begin{figure}[t]
	\centering
	\includegraphics[width=\columnwidth]{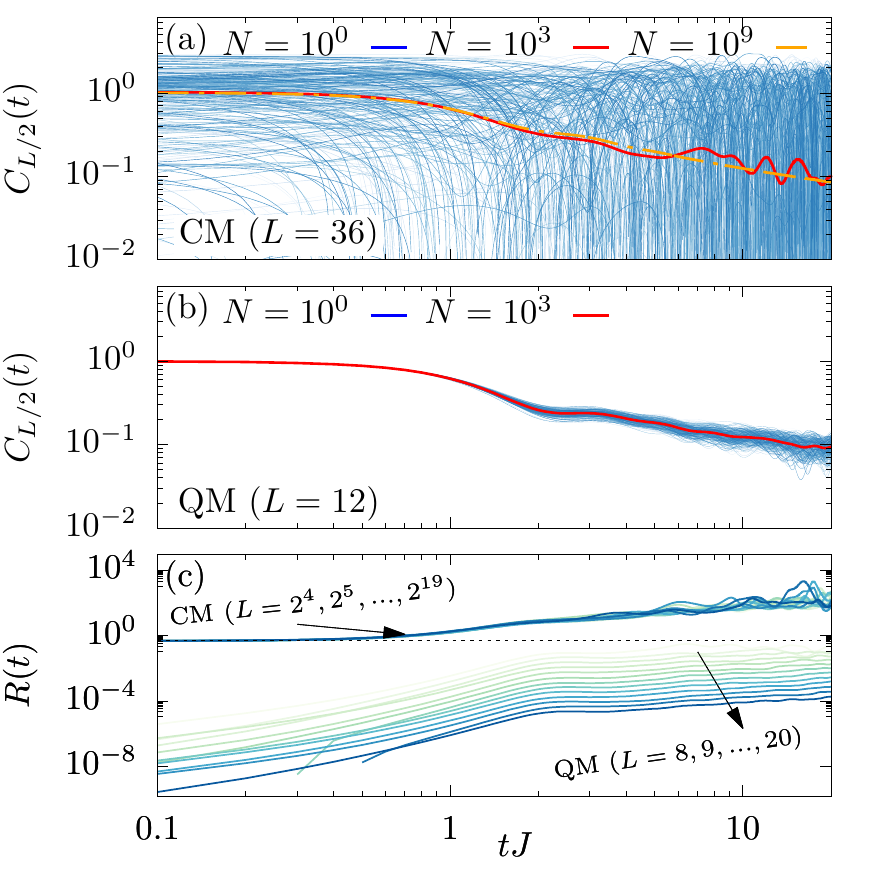}
	\caption{Single-state trajectories of $C_{L/2}(t)$ (blue lines) 
	for \numtrajectories\ random initial states
	in the XXZ chain with $\Delta=1.5$ in the classical case (a)
	and the quantum case (b). Red and orange lines show the corresponding averages
	over $N=10^3$ and $N=10^9$ (only classical) trajectories.
	(c) Relative variance $R(t)$ of sample-to-sample fluctuations
	as obtained by
	Eq.~\eqref{eq:sample-to-sample-fluctuations}
	for $N=10^3$ and different system sizes
	$L=2^4,2^5,\dots,2^{19}$ (classical) and $L=8,9,\dots,20$ (quantum).
	Dashed line indicates the value of $R(0)$ for the classical case which is
	essentially the second moment of the probability distribution for
	$\anylocaldensity_{L/2}^2(0)$ that arises from the random initial
	configuration of the state.
	Note that we consider the high-temperature limit $\beta\rightarrow0$ 
	and that $C_{L/2}(0)$ is set to $1$.
	}
	\label{fig:typicality-vs-classicalaveraging}
\end{figure}
The simulation of the classical spin systems is done by numerically solving the
Hamiltonian equations of motion \eqref{eq:hamiltonian-eom} using a fourth-order
Runge-Kutta (RK4) scheme.
We use a time step $\delta t$ that is small enough to ensure
that the total energy and magnetization are conserved to very high accuracy.
The computational complexity of the simulation of classical systems
growths only linearly in their
system size $L$ and is mainly determined by the number $N$ of samples
used in the averaging \eqref{eq:classical-averaging}.
Importantly, there exists no analog of typicality in classical mechanics,
such that we have to average over many samples $N \gg 1$ of independent random
initial-state configurations, no matter how large the system size $L$.
This crucial difference between classical and quantum simulations is
illustrated in \figref{fig:typicality-vs-classicalaveraging}, which
shows the single-state trajectories of the equal-site correlation function
$C_{L/2}(t)$ for \numtrajectories\
random initial states
in the classical [cf.\ \figref{fig:typicality-vs-classicalaveraging}(a)]
and the quantum [cf.\ \figref{fig:typicality-vs-classicalaveraging}(b)]
version of the XXZ chain with anisotropy $\Delta=1.5$.
In the classical case, each individual trajectory appears random and
the behavior of $C_{L/2}(t)$ can only be inferred from the average,
whereby the average over $N=10^3$ trajectories still shows significant deviations
from the average over $N=10^9$ trajectories.
In contrast, in the quantum case, the individual random realizations show only
small deviations from the average over $N=10^3$ states, even for the small
system size $L=12$ used here.
\figref{fig:typicality-vs-classicalaveraging}(c) shows the
corresponding relative variance of the sample-to-sample fluctuations,
\begin{align}\label{eq:sample-to-sample-fluctuations}
	R(t)=\frac{\overline{C_{L/2}(t)^2}-\overline{C_{L/2}(t)}^2}{\overline{C_{L/2}(t)}^2}\ ,
\end{align}
for different system sizes $L=2^4,2^5,\dots,2^{19}$ (in the classical case)
and $L=8,9,\dots,20$ (in the quantum case).
Here, the overbar
in
$\overline{C_{L/2}(t)}$
denotes the average over $N=10^3$ samples.
Note that a
given quantity, here $C_{L/2}(t)$, is sometimes called self-averaging if $R(t)$
decreases with increasing $L$, e.g., $R(t) \propto L^{-1}$ is referred to as
strong self-averaging in Ref.~\cite{Schiulaz2020}.
As shown in \figref{fig:typicality-vs-classicalaveraging}(c),
in the classical
case, $R(t=0)$ starts at a value that results from the second
moment of the probability distribution for the initial value
$\anylocaldensity_{L/2}^2(0)$.
We find that $R(t)$ increases with time, as the average $\overline{C_{L/2}(t)}$ itself decays
to smaller and smaller values.
Importantly, there is no dependence on system size in this behavior, even for
the exponentially increasing $L$.
Thus, self-averaging is
absent in the case of classical dynamics such that large values of $N$ are
necessary to faithfully capture the ensemble average also for large system
sizes $L$.

In the quantum case, $R(t)$ shows a similar increase in time as above,
while at the same time being orders of magnitudes smaller than in the
classical case -- even for the smallest system size $L=8$ shown.
Crucially, for increasing system size, $R(t)$ decreases exponentially in line
with the typicality estimate \eqref{eq:typicality-error}.
In
this sense, quantum typicality can be seen as an extreme form of self-averaging
as exponentially less random realizations are required at larger $L$ to
accurately determine the full ensemble average.

\section{Results}\label{sec:results}
The discussion of our numerical results includes the comparison between
quantum and classical density dynamics of magnetization in the
1D XXZ chain in \sref{sec:results-XXZ-chain},
magnetization and energy in the
quasi-1D XXX ladder in \sref{sec:results-XXX-ladder},
as well as charge in the Fermi-Hubbard model
in \sref{sec:results-hubbard}.
In all cases, we will focus on time scales where the bulk of the density
distribution is still reasonably concentrated around the center and away
from the boundaries.
The time dependence of the classical correlation functions is always rescaled
by the factor $\tilde{s}=\sqrt{s(s+1)}$ to account for the different
lengths of quantum and classical spins. For the quantum spin $s=1/2$
considered here, this factor is $\tilde{s}\approx0.87$.

\subsection{Magnetization dynamics in the 1D XXZ chain}
\label{sec:results-XXZ-chain}

\begin{figure}[tb]
	\centering
	\includegraphics[width=\columnwidth]{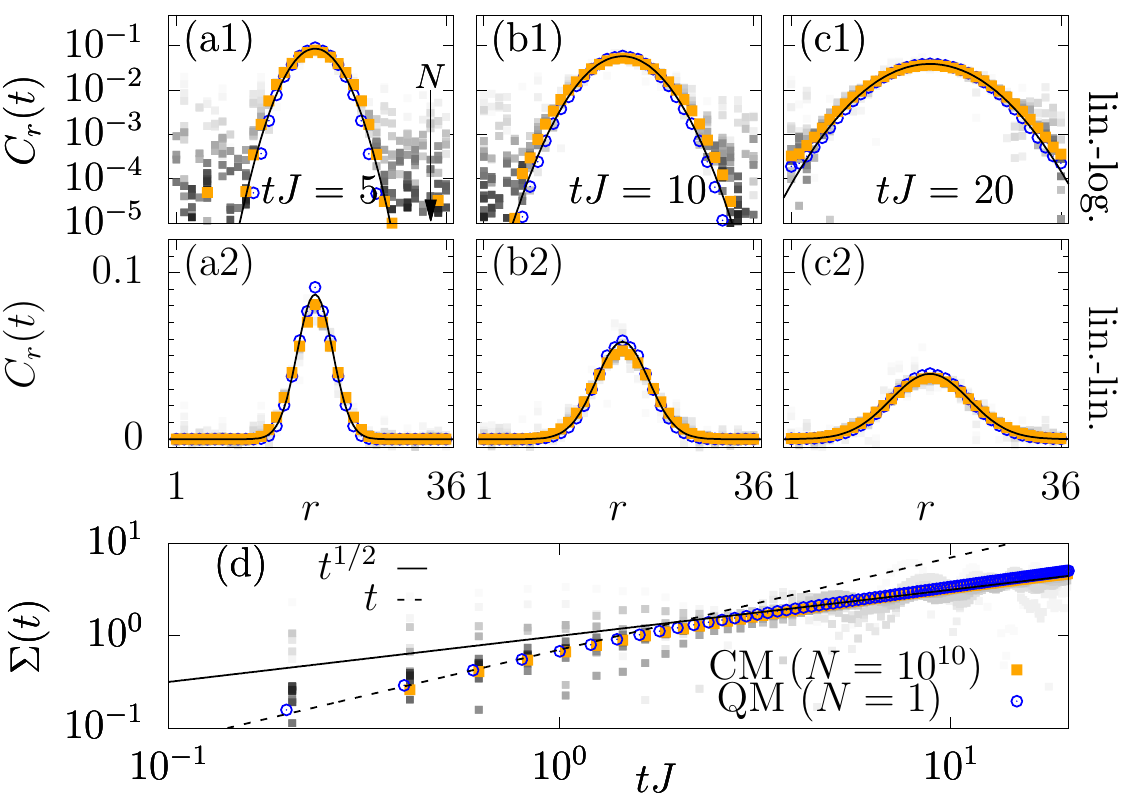}
	\caption{
		(a1)-(c2) Profiles $C_r(t)$ of magnetization densities in the XXZ spin chain
		\eqref{eq:local-hamilton-spinchain} with anisotropy $\Delta=1.5$
		and $L=36$, sampled over $N$ initial states.
		Solid lines are Gaussian fits \eqref{eq:gaussianfit} to the QM data.
		(d) Time-dependent spatial width $\Sigma(t)$ as obtained by
		Eq.~\eqref{eq:spatial-variance}.
		Dashed and solid lines indicate scaling $t^{\alpha}$ for $\alpha=1$
		and $1/2$.
		To illustrate the necessity for large sample sizes in the simulations of classical dynamics,
		additional CM data is shown for different sample sizes $N=2^l\cdot10^3$
		with $l=0,1,\dots,20$ (grey scales).
		Note that we consider the high-temperature limit $\beta\rightarrow0$ and that the legend in panel (d) applies to all other panels.
		}
	\label{fig:chain_D15_samples}
\end{figure}
\begin{figure}[b]
	\centering
	\includegraphics[width=\columnwidth]{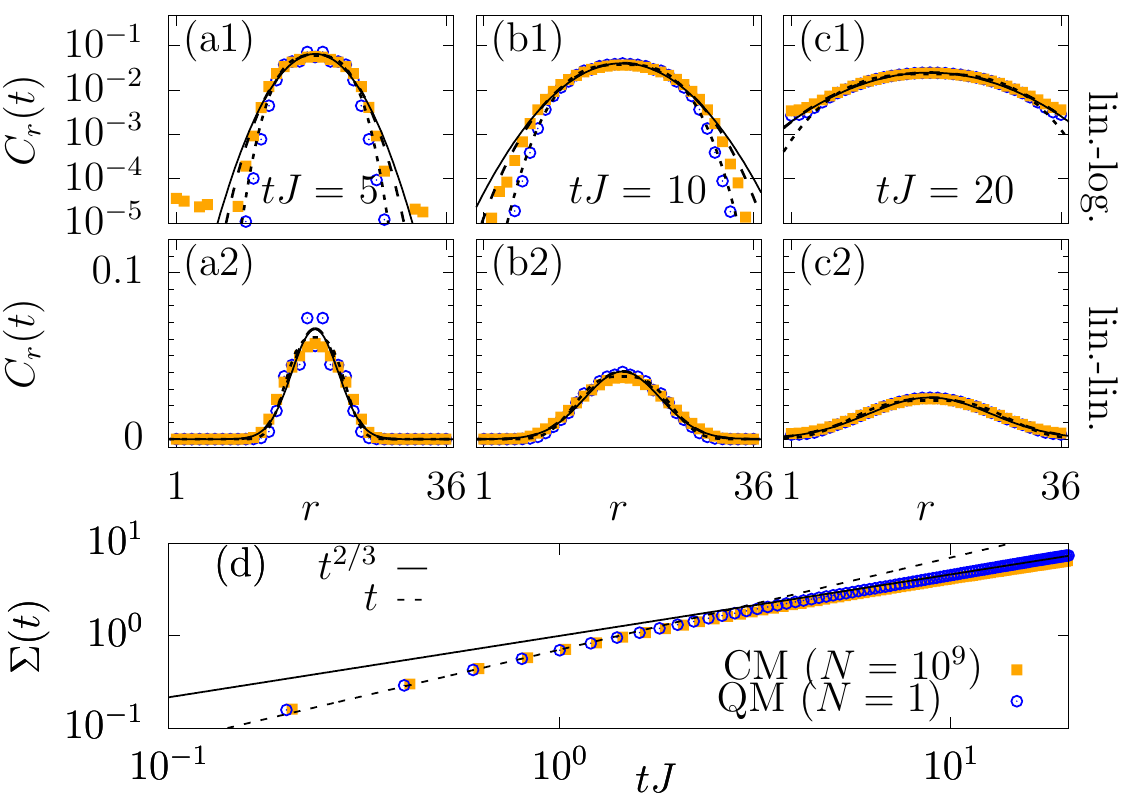}
	\caption{(a1)-(c2) Profiles $C_r(t)$ of magnetization densities in the XXZ spin chain
		\eqref{eq:local-hamilton-spinchain} with anisotropy $\Delta=1.0$
		and $L=36$, sampled over $N$ initial states.
		Solid lines are Gaussian fits \eqref{eq:gaussianfit} to the QM data.
		Dashed lines indicate KPZ scaling functions \cite{Prahofer2004}.
		Dotted lines indicate a function ${\propto\exp(-a\vert r-L/2\vert^3)}$.
		(d) Time-dependent spatial width $\Sigma(t)$ as obtained by
		Eq.~\eqref{eq:spatial-variance}.
		Dashed and solid lines indicate scaling $t^{\alpha}$ for $\alpha=1$
		and $2/3$.
		Note that we consider the high-temperature limit $\beta\rightarrow0$ and that the legend in panel (d) applies to all other panels.
		}
	\label{fig:chain-D10}
\end{figure}
\begin{figure}[h!]
	\centering
	\includegraphics[width=\columnwidth]{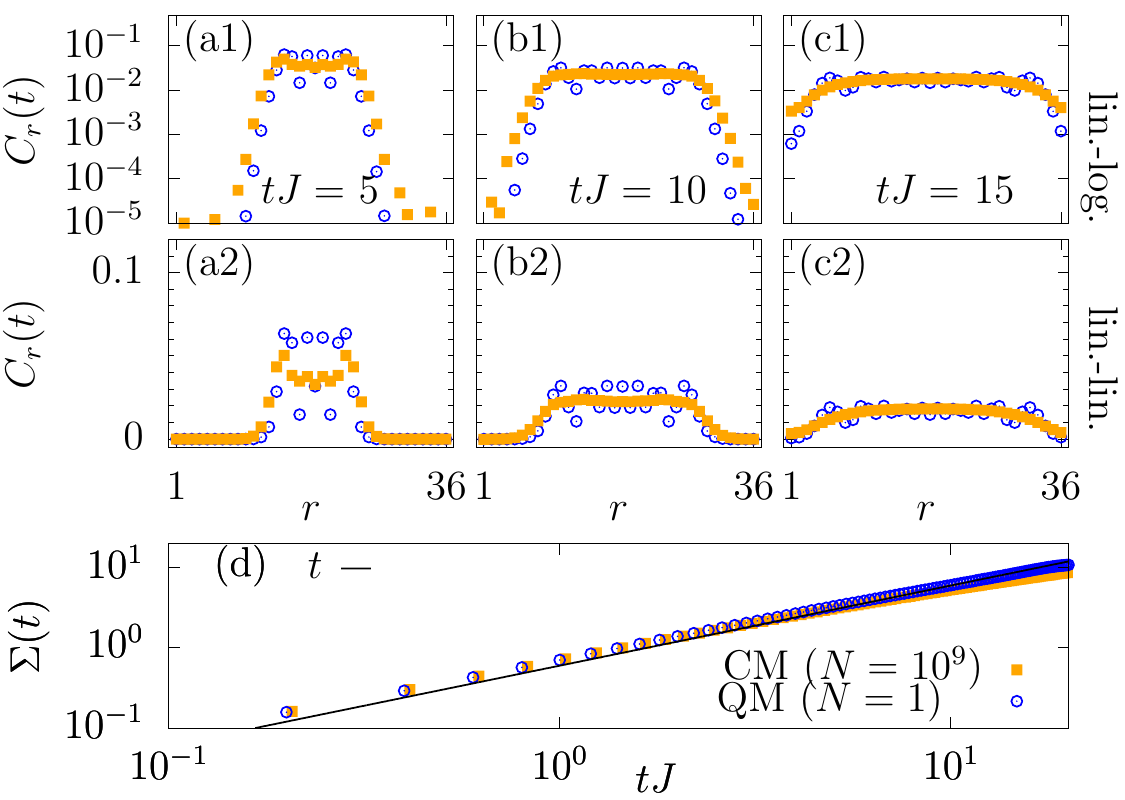}
	\caption{(a1)-(c2) Profiles $C_r(t)$ of magnetization densities in the XXZ spin chain
		\eqref{eq:local-hamilton-spinchain} with anisotropy $\Delta=0.5$
		and $L=36$, sampled over $N$ initial states.
		(d) Time-dependent spatial width $\Sigma(t)$ as obtained by
		Eq.~\eqref{eq:spatial-variance}.
		Solid line indicates scaling $t^{\alpha}$ for $\alpha=1$.
		Note that we consider the high-temperature limit $\beta\rightarrow0$ and that the legend in panel (d) applies to all other panels.
		}
	\label{fig:chain-D05}
\end{figure}

We first focus on the dynamics of magnetization in the integrable 1D XXZ chain
\eqref{eq:local-hamilton-spinchain} of size $L=36$ for different values of the
anisotropy $\Delta=0.5,1,$ and $1.5$.

Starting with the anisotropy $\Delta=1.5$,
\figref{fig:chain_D15_samples}{(a1)-(c2)} shows the corresponding profiles
$C_{r}(t)$ from quantum and classical dynamics
at fixed times $t$ in linear
and semilogarithmic plots. For all values of $t$, the quantum and the
classical profiles show a very good agreement and are accurately described by
Gaussian functions \eqref{eq:gaussianfit} that broaden over time. This is in
line with the diffusive transport that is expected in the regime $\Delta>1$
\cite{Bertini2021}.
In order to illustrate the necessity for extensive averaging of the classical
data, we show additional data for different sample sizes $N=2^l\cdot10^3$ with
$l=0,1,\dots,20$ in the same plots. While the time dependence of $C_{r}(t)$ in
the center of the chain is already reasonably well captured for smaller sample
sizes $N=\mathcal{O}(10^3)$, the level of noise away from the center is
considerable and only sufficiently suppressed for the largest sample sizes
$N=\mathcal{O}(10^9)-\mathcal{O}(10^{10})$.
In the following, we will thus always use a rather large sample size
$N=10^9$ for our simulations of classical systems.

In addition to the space-time profiles, \figref{fig:chain_D15_samples}(d)
shows the time dependence of the corresponding spatial width $\Sigma(t)$ as
obtained by Eq.~\eqref{eq:spatial-variance}.
Naturally, an accurate calculation of $\Sigma(t)$ also relies on a good
signal-to-noise ratio, which is again illustrated by additional data for
smaller sample sizes $N$ in the classical results.
For the largest sample sizes, we see a very good agreement between the quantum
and the classical results, both in the initial ballistic scaling
$\Sigma(t)\propto t $ as well as in the diffusive scaling
$\Sigma(t)\propto \sqrt{t}$ for later times.
The initial
ballistic scaling at short times can be understood as a local expansion of
the spin excitation below its mean free path. Above this mean free path,
the essentially classical hydrodynamic description applies and the
ballistic behavior crosses over to the asymptotic diffusive transport.

Moving on to the isotropic spin chain, \figref{fig:chain-D10} shows
analogous data as above, but now for $\Delta=1.0$. Similarly as before, we
see a good agreement between the classical and the quantum results, albeit
with some small but visible deviations in the profiles $C_{r}(t)$
[cf.\ \figref{fig:chain-D10}(a1)-(c2)].
Especially in the tails of the distributions, we find that the overall shape
of the profiles is no longer described by Gaussian functions
\eqref{eq:gaussianfit}, indicating the shift from normal to anomalous
diffusion. Indeed, the existence of superdiffusion at the isotropic point is
well established
(see Ref.~\cite{Bulchandani2021} and references therein).
More specifically, it has been shown that $C_r(t)$ is well
described by the KPZ scaling function, which is similar to a Gaussian
in the bulk of the distribution, but exhibits faster than Gaussian decay in the
tails.
Yet, for the system sizes and times shown here, the agreement with the KPZ scaling function
is not fully developed, and the data is rather described by a function ${\propto\exp(-a\vert r-L/2\vert^3)}$.
The anomalous transport
is also reflected in the scaling of the spatial width
$\Sigma\propto t^{\alpha}$ with $\alpha=2/3$, which is captured both by the
quantum and the classical dynamics [cf.\ \figref{fig:chain-D10}(d)].

As the final comparison in the 1D XXZ chain, \figref{fig:chain-D05} shows
data for anisotropy $\Delta=0.5$. In this regime, the quantum dynamics is
dominated by an extensive set of conservation laws and a good agreement
between quantum and classical dynamics can no longer be expected.
This expectation is confirmed by the space-time profiles $C_{r}(t)$
[cf.\ \figref{fig:chain-D05}(a1)-(c2)], where we observe noticeable differences
between the classical and the quantum results.
Interestingly, however, the rough shape of the
profiles as well as the overall speed at which they spread over time are
captured quite well by the classical results -- at least on the time scales
shown here. This also pertains to the scaling of the spatial width,
$\Sigma(t)\propto t$ [cf.\ \figref{fig:chain-D05}(d)],
which indicates the ballistic transport that has been
rigorously proven to exist for the quantum system in the thermodynamic limit
\cite{Prosen2011,Prosen2013,Ilievski2016}. However, for longer times
$tJ\gtrsim 10$, a slowdown in the scaling of the width $\Sigma(t)$ in the
classical data becomes noticeable.

The similarities between the quantum and the classical data in
\figref{fig:chain-D05} might indicate that taking the classical limit
$s\to\infty$ appears to be a rather weak form of integrability breaking. In
particular, while the quantum $s = 1/2$ model features strict ballistic
transport, the classical model is expected to be fully chaotic and therefore
to exhibit diffusive transport at $\Delta = 0.5$ (especially since we are now
away from the potentially special point $\Delta = 1$). However, as becomes
clear from \figref{fig:chain-D05}, this conjectured diffusive behavior in
the classical chain must set in at significantly longer time and length
scales.
For some additional data from classical dynamics in a larger system
of size $L=2000$, see \aref{app:longXXZ}.

\subsection{Magnetization and energy dynamics in the quasi-1D XXX ladder}
\label{sec:results-XXX-ladder}
Next, we move from 1D chains to quasi-1D spin ladders
\eqref{eq:local-hamilton-spinladder}, where the integrability of the quantum
system is broken. We compare the quantum and classical dynamics for magnetization and energy
in an isotropic spin ladder of length $L=20$.
Note that this corresponds to $40$ spin-$1/2$ lattice sites in
total, which is far beyond the range of standard exact diagonalization and
close the maximum system sizes that are nowadays in reach of
massively parallelized simulations on
state-of-the-art supercomputing clusters.
The transport of both magnetization and energy in the quantum
case $s=1/2$ is known to be diffusive in this model
\cite{Richter2019f}.

\begin{figure}[tb]
	\centering
	\includegraphics[width=\columnwidth]{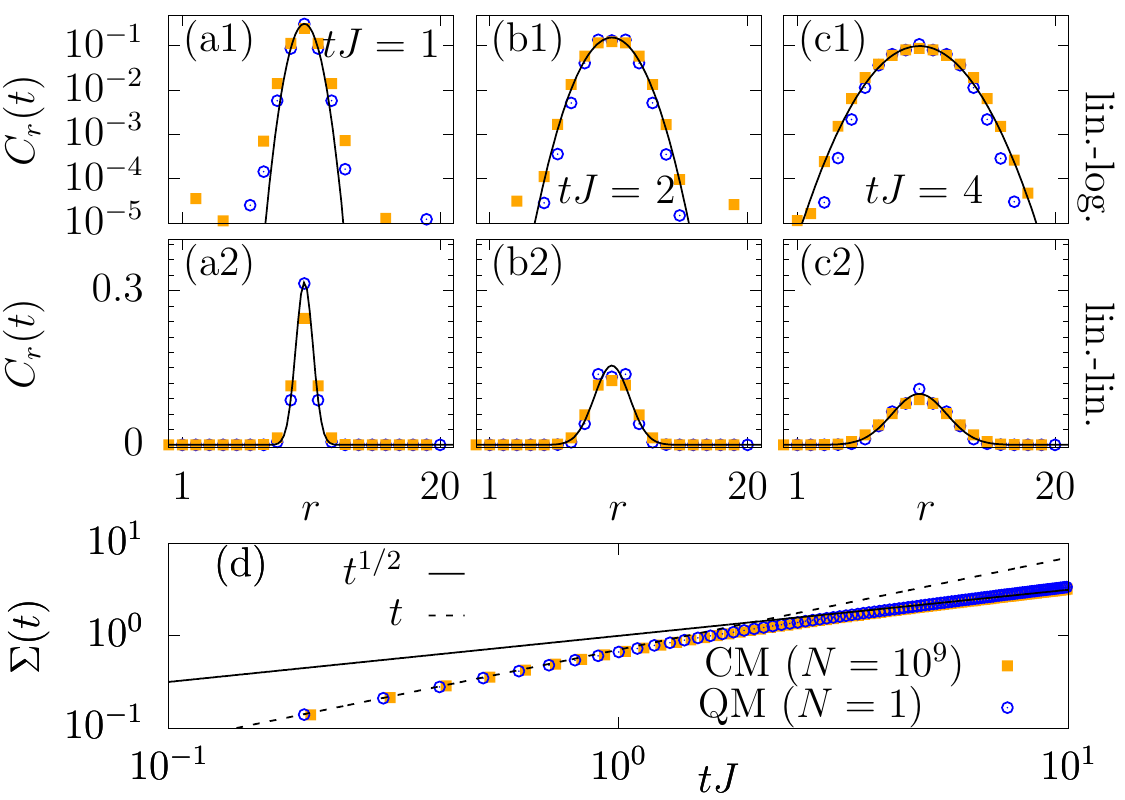}
	\caption{(a1)-(c2) Profiles $C_r(t)$ of magnetization densities in the
		XXX spin ladder
		\eqref{eq:local-hamilton-spinladder} with $L=20$, sampled over $N$
		initial states.
		Solid lines are Gaussian fits \eqref{eq:gaussianfit} to the QM data.
		(d) Time-dependent spatial width $\Sigma(t)$ as obtained by
		Eq.~\eqref{eq:spatial-variance}.
		Dashed and solid lines indicate scaling $t^{\alpha}$ for $\alpha=1$
		and $1/2$.
		Note that we consider the high-temperature limit $\beta\rightarrow0$ and that the legend in panel (d) applies to all other panels.
		}
	\label{fig:ladder-mag}
\end{figure}

\figref{fig:ladder-mag} shows the space-time profiles $C_{r}(t)$ and the
spatial width $\Sigma(t)$ for magnetization.
Again, the profiles $C_{r}(t)$ show a very good agreement
in the comparison between the quantum and the classical results and are
accurately described by Gaussian functions
\eqref{eq:gaussianfit}. Additionally, the corresponding spatial width
$\Sigma(t)$ agrees very well and the quantum and classical results lie
on top of each other, from the initial times of ballistic scaling
$\Sigma\propto t$ up to later times of diffusive scaling
$\Sigma(t)\propto\sqrt{t}$.

\begin{figure}[b]
	\centering
	\includegraphics[width=\columnwidth]{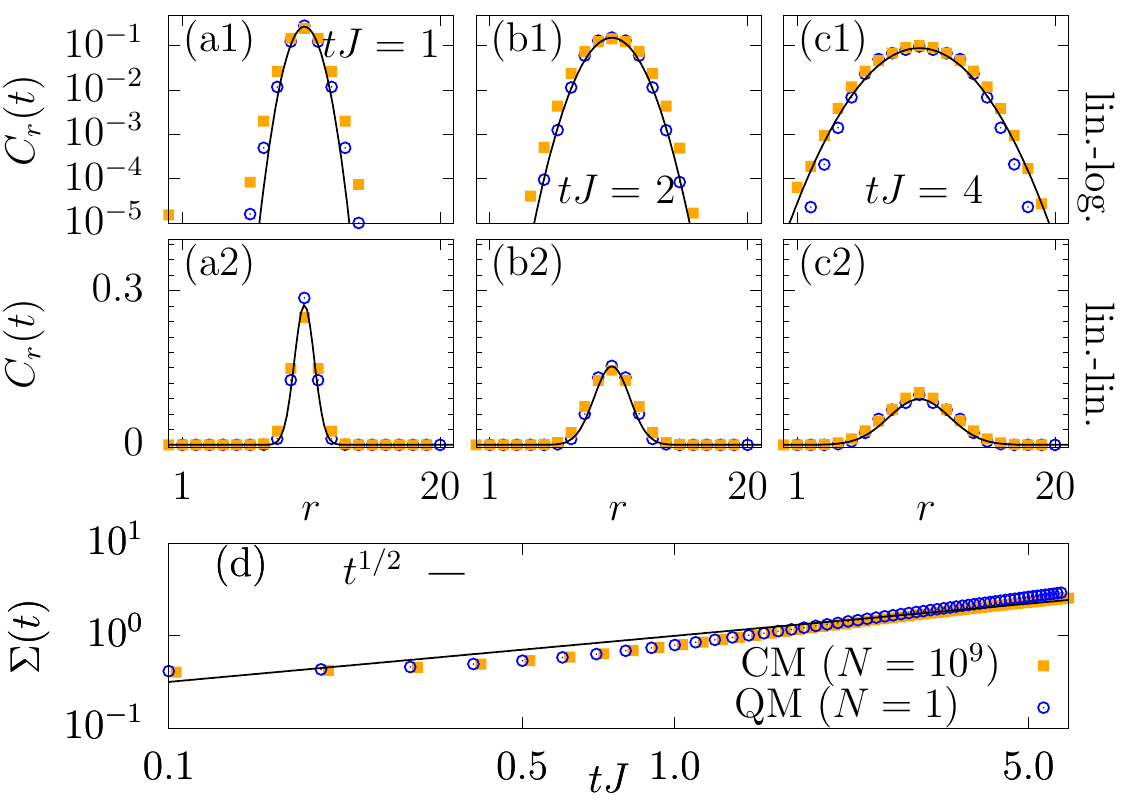}
	\caption{(a1)-(c2) Profiles $C_r(t)$ of energy densities in the XXX spin ladder
		\eqref{eq:local-hamilton-spinladder} with $L=20$, sampled over $N$
		initial states.
		Solid lines are Gaussian fits \eqref{eq:gaussianfit} to the QM data.
		(d) Time-dependent spatial width $\Sigma(t)$ as obtained by
		Eq.~\eqref{eq:spatial-variance}.
		Solid line indicates scaling $t^{\alpha}$ for $\alpha=1/2$.
		Note that we consider the high-temperature limit $\beta\rightarrow0$ and that the legend in panel (d) applies to all other panels.
		}
	\label{fig:ladder-energy}
\end{figure}

\figref{fig:ladder-energy} shows the same data as
\figref{fig:ladder-mag}, but for the dynamics of local energy. The quantum
and classical results again agree very well and match the typical signatures
of diffusive transport. The only difference compared to the results in
\figref{fig:ladder-mag} lies in the initial scaling
of the spatial width $\Sigma(t)$, which, owing to the broader initial peak for
local energy densities, does start at a nonzero initial value.


The good agreement between quantum and classical dynamics in
the XXX ladder might not be entirely surprising due to the fact that we are
considering high
temperatures $\beta \to 0$ and that both the quantum and the classical model
are nonintegrable. However, we still find it remarkable that $C_r(t)$ agrees
quantitatively on a very detailed level, leading to an essentially
indistinguishable dynamics of the mean-squared displacement $\Sigma(t)$.
Crucially, the latter is directly related to physically important quantities
such as the diffusion coefficient. Our results in
Figs.~\ref{fig:ladder-mag} and \ref{fig:ladder-energy}
suggest that this diffusion
coefficient is the same in the quantum and the classical model, emphasizing
that (the significantly less costly) simulations of classical systems can
provide
a useful strategy to gain insights into the properties of strongly correlated
quantum many-body systems. While not shown here, we expect a similarly good
agreement between quantum and classical dynamics also for XXZ spin ladders
	[i.e., when incorporating an anisotropy $\Delta$ in the Hamiltonian
		\eqref{eq:local-hamilton-spinladder}]. In particular, studying the equal-site
correlation functions $C_{L/2}(t)$ of magnetization and
energy, \cite{Schubert2021} found a convincing agreement between the quantum
and classical dynamics in ladders with $\Delta=0.5,1,$ and $1.5$.


\subsection{Charge dynamics in the Fermi-Hubbard chain}
\label{sec:results-hubbard}
\begin{figure}[tb]
	\centering
	\includegraphics[width=\columnwidth]{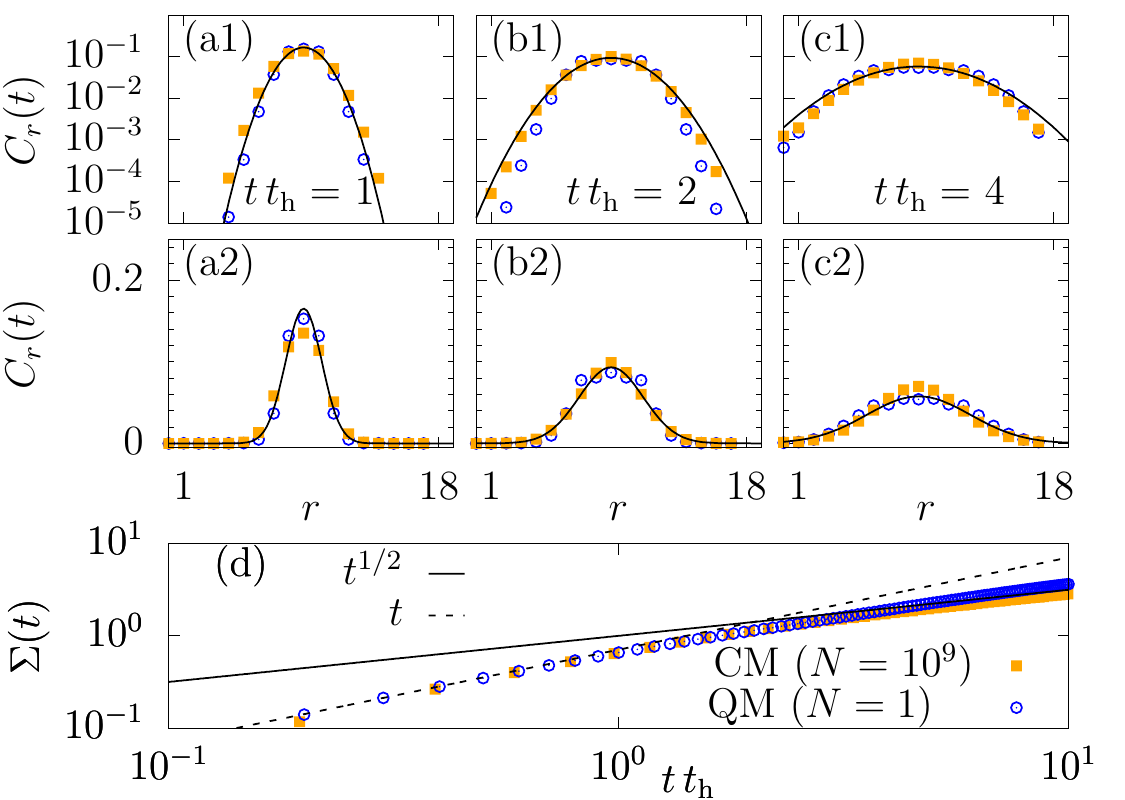}
	\caption{(a1)-(c2) Profiles $C_r(t)$ of charge densities in the Fermi-Hubbard chain
		\eqref{eq:local-hamilton-hubbard} with $L=18$ and $U/\thubbard=4$, sampled over $N$
		initial states.
		Solid lines are Gaussian fits \eqref{eq:gaussianfit} to the QM data.
		(d) Time-dependent spatial width $\Sigma(t)$ as
		obtained by Eq.~\eqref{eq:spatial-variance}. Dashed and solid lines
		indicate scaling $t^{\alpha}$ for $\alpha=1$ and $1/2$.
		Note that we consider the high-temperature limit $\beta\rightarrow0$ and that the legend in panel (d) applies to all other panels.
		}
	\label{fig:hubbard-u4}
\end{figure}
Finally, we turn to the dynamics of local charge densities in the
integrable Fermi-Hubbard chain, where earlier numerical
studies have found clear signatures of diffusive charge dynamics for strong
interactions $U/t_\mathrm{h} \approx 16$ \cite{Jin2015,Steinigeweg2017}.
However, let us note that
this observation of diffusion is at odds with generalized
hydrodynamics results, which predict the occurrence of superdiffusive charge
transport \cite{Bertini2021}, given the SU$(2)$ symmetry of the Fermi-Hubbard
model (similar to the
case of spin transport in the isotropic spin-$1/2$ Heisenberg chain).
Here, we
consider a somewhat lower interaction strength,
$U/\thubbard=4$, and study chains of length $L=18$. Let us stress again that,
while the Fermi-Hubbard chain has no obvious classical limit, the Jordan-Wigner
transformation in Eq.~\eqref{eq:local-hamilton-hubbard-to-heisenberg} and the
subsequent limit $s \to \infty$ allows for a comparison with classical dynamics
also in this case.

\figref{fig:hubbard-u4} shows the corresponding space-time profiles
$C_{r}(t)$ and the spatial width $\Sigma(t)$.
Comparing the results for the quantum and the classical dynamics, we see
a good agreement in the space-time profiles for all values of
$t$ shown here.
While the profiles are well described by
Gaussians \eqref{eq:gaussianfit} in the bulk of the system, we observe
notable deviations from these Gaussian fits in the tails of the
distributions [cf.\ \figref{fig:hubbard-u4}(b1)]. This might be
reminiscent of the
superdiffusive KPZ scaling of spin transport in the Heisenberg chain discussed
above in \figref{fig:chain-D10}. Then again, the overall
broadening of the profiles seems to follow a diffusive
scaling, $\Sigma(t)\propto\sqrt{t}$
[cf.\ \figref{fig:hubbard-u4}(d)], both for quantum and classical
dynamics. To be more precise, at the longest time $t\thubbard = 10$ shown in
\figref{fig:hubbard-u4}(d), we actually do observe some slight deviations
in the time dependence of $\Sigma(t)$, where the broadening in the classical
case becomes notably slower compared to the quantum case. This observation
might hint at the possibility that the nonintegrable classical model
supports diffusion, while the original integrable Fermi-Hubbard chain
asymptotically shows a crossover to superdiffusion. However, resolving the
latter is numerically quite challenging.

Finally, we note that the remarkable agreement between
quantum and classical dynamics observed for $U/\thubbard = 4$ in
\figref{fig:hubbard-u4} can in general neither be expected for very
small interactions $U/\thubbard\rightarrow 0$
nor for much stronger interactions.
On the one hand, for weak interactions, the charge dynamics becomes more and
more ballistic, as the model is approaching the limit of free fermions.
On the other hand, for much stronger values of $U$, the on-site interaction
dominates the dynamics and reduces the effective number of interacting neighbors
in the system \cite{Elsayed2015}, which is expected to affect the comparability between quantum
and classical dynamics.

\FloatBarrier

\section{Conclusion}
\label{sec:conclusion}
In this paper, we have compared the quantum and classical dynamics of
spatiotemporal density-density correlation functions
in different (quasi-)one-dimensional systems for high temperatures
$T\rightarrow\infty$.
In the quantum case, we employed the concept of quantum typicality
in combination with an efficient forward propagation of pure states
to obtain results in spin-$1/2$ systems with up to $40$ lattice sites
with an extremely low level of statistical noise.
In order to achieve a similar signal-to-noise ratio in the classical case, we
performed extensive averaging over large samples of
$N=\mathcal{O}(10^9)-\mathcal{O}(10^{10})$ classical trajectories.
Based on the comparison of space-time profiles of spin and energy correlations,
we found a remarkably good agreement between quantum and classical
dynamics -- not only in cases where both the quantum and classical model
are nonintegrable, but also in cases where the quantum spin-$1/2$ model is
integrable and the corresponding classical $s\to\infty$ model is not.
Further, we found that this agreement not only holds in the bulk but also in the
tails of the density distributions.
The good agreement between quantum and classical results also manifested itself in
the time dependence of the mean-squared displacement of the
density profiles, which exhibited very similar scaling for quantum and classical
models, at least on the time and length scales considered here.

Furthermore, we showed
that such a correspondence between quantum and classical
dynamics can also be achieved in less obvious cases where the
original quantum system is not directly written in spin language. In
particular, we considered the one-dimensional Fermi-Hubbard model, which by means
of a Jordan-Wigner transform can be brought into the form of a particular type
of spin ladder, for which we then take the $s\to \infty$ limit.
The results from the simulations of quantum and classical dynamics showed a good agreement,
both for the space-time profiles of local charge as well as the time dependence of the
corresponding spatial width, at least for the interaction strength considered here.
This agreement is expected to break down for smaller interaction strengths, where the
Fermi-Hubbard model approaches the integrable limit of free particles, as well as for
much stronger interaction strengths, where the effective number of interacting
neighbors per site is reduced significantly
\cite{Elsayed2015}.

There are several future directions of research to explore. Apart from the question
how far the agreement between quantum and classical dynamics carries over to finite
temperatures, it would also be interesting to further explore the Fermi-Hubbard model
in more detail. For instance, one might expect that the agreement between quantum and
classical dynamics increases in the extended Fermi-Hubbard model, where additional
interactions increase the effective number of interacting neighbors.
Moreover a study in higher spatial dimensions would be interesting, where
the range of the simulation of quantum systems is severely limited.

\section*{Acknowledgments}

This work has been financially supported by the Deutsche
Forschungsgemeinschaft (DFG), Grants No. 397067869 (STE 2243/3-2)
and 397300368 (MI 1772/4-2),
within the DFG Research Unit FOR 2692, Grant No. 355031190.
J.R. has been funded by the European Research Council (ERC) under the
European Union’s Horizon 2020 research and innovation programme
(Grant Agreement No. 853368). Additionally, we gratefully acknowledge
the computing time, granted by the “JARA-HPC Vergabegremium” and
provided on the “JARA-HPC Partition” part of the supercomputer “JUWELS”
at Forschungszentrum Jülich.

\begin{appendix}

	\section{Large classical XXZ chain with $\Delta=0.5$}
	\label{app:longXXZ}

	Complementary to the results for the XXZ chain with anisotropy
	$\Delta=0.5$ discussed in \sref{sec:results-XXZ-chain}, we here present
	additional classical results for longer times and a significantly larger
	chain of size $L=2000$.
	\begin{figure}[t]
		\centering
		\includegraphics[width=\columnwidth]{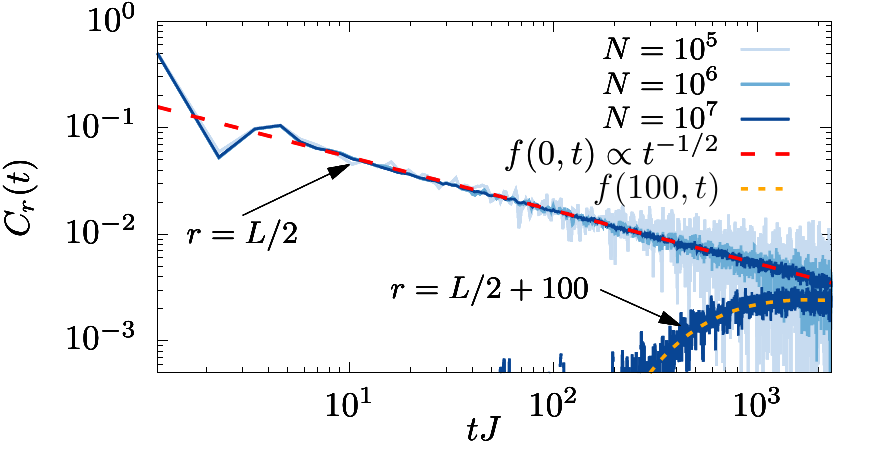}
		\caption{Classical dynamics of $C_r(t)$ for $r=L/2$ and $r=L/2+100$ in a
			XXZ spin chain with anisotropy $\Delta=0.5$ and large system size $L=2000$.
			Dashed lines indicate a Gaussian fit to the equal-site correlation function
			$C_{L/2}(t)$ [cf.\ Eq.~\eqref{eq:simplegauss}].
			Note that we consider the high-temperature limit $\beta\rightarrow0$.
			}
		\label{fig:longXXZ}
	\end{figure}

	\figref{fig:longXXZ} shows the corresponding dynamics of the equal-site
	correlation function $C_{L/2}(t)$ for different sample sizes
	$N=\mathcal{O}(10^5)-\mathcal{O}(10^6)$.
	For the largest sample size shown, a diffusive decay
	$C_{L/2}(t)\propto t^{-\alpha}$
	with $\alpha=1/2$ becomes visible on longer time scales, which
	follows from the Gaussian profile [cf.\ Eq.~\eqref{eq:gaussianfit}]
	\begin{align}\label{eq:simplegauss}
		f(\tilde{r},t)=\frac{1}{\sqrt{4\pi Dt}}
		\exp\left(-\frac{\tilde{r}^{2}}{4 Dt}\right)\ , \quad \tilde{r}=r-L/2
	\end{align}
	where the diffusion constant $D$ may serve as a single fit parameter.
	In addition to the data for $C_{L/2}(t)$, we also show the correlation
	function $C_r(t)$ at a site $r=L/2+100$ far away from the center of
	the chain, where $C_r(t)$ starts in the initial infinite-temperature
	many-body background and increases at times when the density peak in the
	center of the chain has spread sufficiently far over the system.
	Remarkably, despite the considerable fluctuations that are still present
	for the shown sample size, the time dependence of $C_{r}(t)$ appears to
	be well captured by the function \eqref{eq:simplegauss}. Note that we do
	not perform another fit, but instead reuse the diffusion constant $D$
	obtained in the fit to $C_{L/2}(t)$.

	However, a genuine confirmation of diffusion would again require the
	study of the full spatial dependence of the density distributions $C_{r}(t)$.
	This in turn necessitates a substantially larger sample size $N$, which,
	given the combination with large $L$ and long time scales, remains
	numerically challenging.
	A more instructive approach to the transport behavior in larger systems
	might be to study the density dynamics in momentum space, i.e., the decay
	of long-wavelength Fourier modes of the real-space data $C_{r}(t)$, for a
	different class of initial states \cite{Steinigeweg2012}.

\end{appendix}

\bibliographystyle{apsrev4-1_titles}
\bibliography{paper.bib}

\end{document}